\documentclass[%
 reprint,
 amsmath,amssymb,
 aps,
]{revtex4-2}

\usepackage{graphicx}
\usepackage{dcolumn}
\usepackage{bm}
\usepackage{amsthm}
\usepackage{xcolor}
\usepackage{hyperref}

\begin{document}

\preprint{APS/123-QED}

\title{Thermodynamic Constraints on Perfect Equilibrium Superconducting Diodes}

\author{Pavan Hosur}
 \affiliation{Department of Physics and Texas Center for Superconductivity, University of Houston, Houston, TX 77204, USA}
 \email{phosur@uh.edu}

\date{\today}

\begin{abstract}
Superconducting diodes promise dissipationless rectification, yet equilibrium platforms without engineered junctions typically exhibit modest efficiencies. We identify a general thermodynamic origin of this behavior that is largely independent of microscopic details. Denoting $\epsilon = |I_c^-/I_c^+|$, where $I_c^\pm$ are critical currents in opposite directions with $|I_c^+|>|I_c^-|$ by convention, we show that an exact perfect equilibrium diode ($\epsilon=0$) is forbidden by continuity of the free energy. Asymptotically perfect behavior $\epsilon\to0$ is possible within a global equilibrium description, but requires free-energy singularities to enable softening of the unfavorable current-carrying branch. We demonstrate this explicitly in an exactly solvable Ising superconductor model. For smooth single-branch superconductors, standard finite-order polynomial Landau theory yields finite lower bounds on $\epsilon$, while Josephson systems obey analogous bounds set by finite harmonic content of the current-phase relation. The high efficiencies often reported in Josephson platforms arise because such devices commonly operate under phase constraints, metastable switching, or external drive, thereby evading unconstrained equilibrium limits. Thus, our results provide a unified framework for interpreting diode efficiencies across superconducting platforms.
\end{abstract}

\maketitle

Recent years have seen explosive resurgence of interest in superconducting diodes (SDs) \citep{MaSDEreview,NadeemSDEreview,NadeemSDEreviewquantum} -- superconducting devices exhibiting unequal critical current magnitudes $|I_{c}^\pm|$ in opposite
directions. They offer enticing technological prospects such as dissipationless rectification and nonreciprocal signal processing, and are natural building blocks for superconducting logic, memory, and neuromorphic elements. As superconducting electronics moves beyond SQUIDs and conventional Josephson-junction (JJ) circuits, understanding which platforms can host highly efficient, scalable SDs has become a key design problem.

A SD response generically appears once time-reversal, inversion, and any spatial symmetry exchanging ``left'' and ``right'' are broken. Producing a nonzero effect is therefore relatively easy: irregular sample geometry, stray vortices, local inhomogeneity, or JJ asymmetry can all generate observable nonreciprocity. As a result, the SD effect has been observed in a myriad of platforms ranging from Van der Waals heterostructures and twisted graphene layers to asymmetric JJs and multiterminal devices \citep{Ando:2019aa,Jiang1994,Ando:2020td,Lyu:2021wg,Du2023,Sundaresh:2023aa,Kealhofer2023,Bauriedl:2022we,Lin:2022aa,Hou2023,castellani2025superconductingfullwavebridgerectifier,Ingla-Aynes2024,Chen:2024aa,Gupta:2023aa,Baumgartner_2022,Paolucci2023,Pal:2022tm,Golod:2022ta,Schmid2025,Jeon:2022aa,Baumgartner:2022wr,Ghosh:2024aa,Banerjee2023,Kim2024,Turini:2022aa,Wu:2022wq,Diez-Merida2021,Chiles:2023aa,Zhang2024,Yun2023,Jeon:2022aa,Anwar:2023aa,Narita:2022tb,Gutfreund:2023aa,Zhao2023,Trahms:2023aa,Yasuda:2019wb,Masuko:2022aa,he2024observationsuperconductingdiodeeffect,Anh:2024aa,shin2024electriccontrolpolarityspinorbit,Margineda:2025aa,Margineda:2023aa,Borgongino:2025aa,Coraiola:2024aa}.
In parallel, it has inspired a wealth of theoretical activity aimed at explaining, enhancing and harnessing the SD effect in various platforms \citep{Yuan2022,Daido2022,Daido2022a,Zhang2022,Davydova2022,Chen2024,Wang2022,He_2022,Zhai2022,MIsaki2021,Ilic2022,Scammell_2022,Zinkl2022,He:2023aa,Jiang2022,Kokkeler2022,Debnath2024,Chazono2023,Vodolazov2005,dePicoli2023,Kochan2023,Ikeda2022,Tanaka2022,Haenel2022,Legg2023,Cuozzo2024,Souto2022,Souto2024,Cheng2023,Steiner2023,Costa2023,Wei2022,Legg2022,Karabassov2022,Hu2023,Wu2023,Nunchot2024,Daido2023,Cayao2024,Banerjee2024,Hosur2023,ChakrabortyBlackSchaffer2025,Soori2024,Soori_2024,Chen2024,Bozkurt2023,Shaffer2024,Lupo2024,fukaya2024designsupercurrentdiodevortex,chen2025enhancedsuperconductingdiodeeffect}. However, these theoretical works are invariably associated with specific types of platforms, and the diversity of platforms and mechanisms obscures the simpler organizing question: when can a superconducting platform exhibit a \emph{large} SD effect rather than merely a finite one?


Assuming without loss of generality that $|I_c^+| > |I_c^-|$, the quality of the SD can be quantified via the diode asymmetry
\begin{equation}
\label{eq:eps-def}
    \epsilon = \left|\frac{I_c^-}{I_c^+}\right| \in [0,1],
\end{equation}
so that a perfect SD corresponds to $\epsilon = 0$. A common figure of merit, the diode efficiency, is defined as $\eta = (1-\epsilon)/(1+\epsilon)$, so $\epsilon \ll 1$ is equivalent to $\eta \approx 1$. Platforms without JJs usually report $\eta \sim 10^{-2}$. A few experiments have achieved $\eta\gtrsim 0.3$ \citep{Bauriedl:2022we,Lin:2022aa,Hou2023,castellani2025superconductingfullwavebridgerectifier,Ingla-Aynes2024}, while a prominent platform, twisted trilayer graphene, reported $\eta\approx1$ at certain fillings \cite{Lin:2022aa}. JJ-based platforms have typically shown higher efficiencies \cite{Paolucci2023,Pal:2022tm,Golod:2022ta,Schmid2025,Jeon:2022aa,Baumgartner:2022wr,Ghosh:2024aa} with at least two realizations achieving $\eta\approx1$ \cite{Chiles:2023aa,Borgongino:2025aa}. Nevertheless, achieving $\epsilon \ll 1$ or $\eta \approx 1$ systematically -- conversely, identifying platform-agnostic principles that hinder high efficiency -- remains a central goal of the field.

In this work, we identify constraints on diode efficiency from the thermodynamics of superconductors. Our central result is that general equilibrium principles impose stringent limits on it. In particular, near-perfect behavior $\epsilon\to0$ requires singularities in the free-energy landscape, such as branch-switching between competing superconducting states, that enable softening of the unfavorable current-carrying branch. Exact $\epsilon=0$ is even more restrictive: it would require a discontinuity of the free energy, which is excluded in conventional local solids. These results explain why $\eta\ll1$ or $\epsilon=O(1)$ is common in unconstrained bulk superconductors, while better efficiencies arise more naturally in Josephson platforms that operate under phase constraints, metastable switching, circuit embedding, or external drive. They also suggest that unusually small $\epsilon$ in a nominally equilibrium platform can serve as a diagnostic of nearby internal structure within the superconducting phase.

\textit{Superconducting thermodynamics setup:}
We consider quasi-1D superconducting systems of length $L$ whose transport is governed by a global phase variable $\Theta$, while additional internal condensate degrees of freedom may relax at fixed $\Theta$. This setting includes multiband or multicomponent superconductors, finite-momentum paired states, and Josephson devices treated within an effectively one-dimensional transport geometry \cite{kruchinin2016multiband,BABAEV201720,yerin2023multipleq,Mizushima:2014aa,Takahashi2014}. Denoting the condensates by $\Delta_j(x)=|\Delta_j(x)|e^{i\phi_j(x)}$ with $j=1,\dots,N$, we separate an overall phase mode $\phi_1(x)\equiv\theta(x)$ from relative phases $\phi_{j>1}(x)=\theta(x)+\alpha_j(x)$. Electromagnetic gauge fields couple only to the overall phase gradient, $\partial_x\theta(x)\rightarrow \partial_x\theta(x)-2eA(x)$,
while the relative phases $\alpha_j$ remain gauge neutral.

For current-biased transport, the appropriate thermodynamic potential is the Gibbs free energy \cite{tinkham2004introduction,McCumber1968},
\begin{equation}
\mathcal{G}
=\mathcal{F}-\frac{I}{2e}\int_0^L(\partial_x\theta+A)\,dx
=\mathcal{F}-\frac{I}{2e}\left(\Theta+\int_0^L A\,dx\right),
\end{equation}
where $\mathcal{F}[\Delta_1(x),\dots,\Delta_N(x)]$ is the Helmholtz free-energy functional relative to the normal state and $\Theta \equiv \int_0^L \partial_x\theta(x)dx$. In Josephson systems, $\Theta$ is the phase difference across the junction, whereas in JJ-free systems, it is the global phase twist across the sample. For a single-component superconductor, $q=\Theta/L$ reduces to the usual Cooper-pair momentum.

At fixed $\Theta$, the equilibrium superconducting state is obtained by minimizing over all internal variables other than the global phase mode:
\begin{equation}
\label{eq:F-eff}
F(\Theta)=
\min_{\{|\Delta_{j\ge1}|,\alpha_{j\ge2}\}}
\mathcal{F}\bigl[\Theta;|\Delta_{1\dots N}|,\alpha_{2\dots N}\bigr].
\end{equation}
The resulting effective free energy $F(\Theta)$ is generally piecewise analytic, since the optimal condensate configuration may change as $\Theta$ is varied. As shown below, $\epsilon\ll1$ requires such singularities while a single smooth branch yields finite lower bounds on $\epsilon$.

Superconductivity is stabilized when $F(\Theta)<0$ relative to the normal state. For notational simplicity, we consider a single connected interval $\Theta\in(\Theta_c^-,\Theta_c^+)$ where superconductivity persists; extension to reentrant cases with multiple intervals is straightforward. In Josephson devices and other mesoscopic settings with flux quantization, $F(\Theta)$ is periodic by $2\pi$ (or $4\pi$ in settings with Majorana zero modes), and we may choose $\Theta_c^\pm=\pm\pi$ (or $\pm2\pi$). In either case, we must have
\begin{equation}
\label{eq:Fqc}
F(\Theta_c^+)=F(\Theta_c^-)
\end{equation}
which equals zero in JJ-free systems and equals the same finite value in Josephson devices.

Within the superconducting domain, the supercurrent and critical currents are defined by
\begin{align}
    I(\Theta)&=
-\partial_A F(\Theta-2eA)\big|_{A=0}
=2e\,\partial_\Theta F(\Theta)
\label{eq:Iq}\\
\label{eq:jc}
I_c^\pm&=\max_\Theta[\pm I(\Theta)]
\end{align}

Their ratio gives the diode asymmetry $\epsilon$ in Eq.~\eqref{eq:eps-def}.

In practical current-biased experiments, switching occurs dynamically when the applied current exceeds the extrema in Eq.~\eqref{eq:jc}. These critical currents are typically reached at $\Theta$ values inside the superconducting interval $(\Theta_c^-,\Theta_c^+)$ rather than at its endpoints. By contrast, $\Theta\notin(\Theta_c^-,\Theta_c^+)$ corresponds to thermodynamic loss of superconducting stability. The constraints derived below rely on the interplay between these dynamic ($I$-based) and thermodynamic ($\Theta$-based) stability criteria.

\textit{Constraints on $\epsilon$:}
We first show that exact $\epsilon=0$ is impossible; any perfect diode limit must therefore be asymptotic. We next show that this limit cannot arise from a single smooth analytic branch of $F(\Theta)$, and instead requires singularities in the free-energy landscape.

\begin{figure}
\includegraphics[width=0.45\columnwidth]{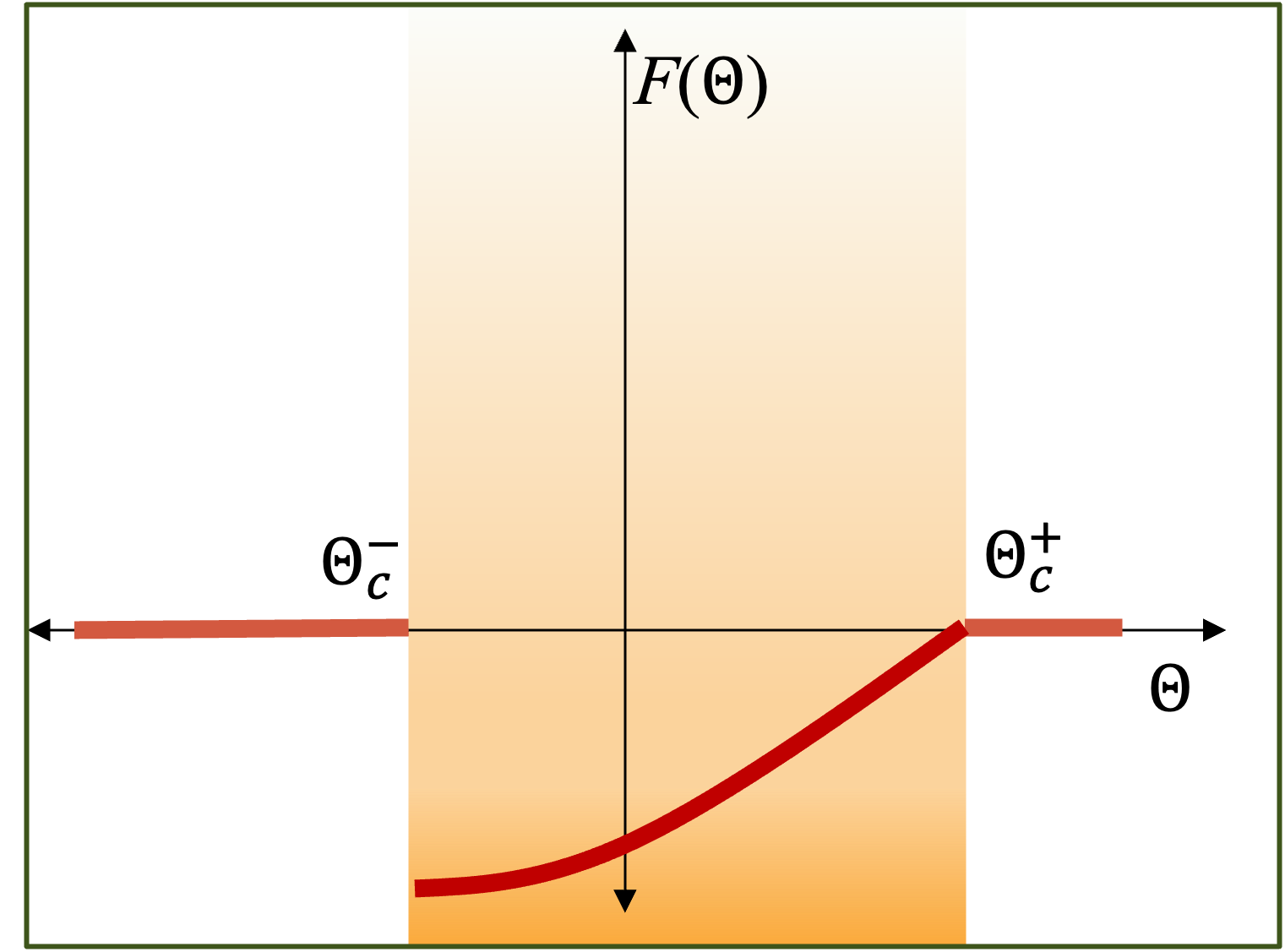}
\includegraphics[width=0.45\columnwidth]{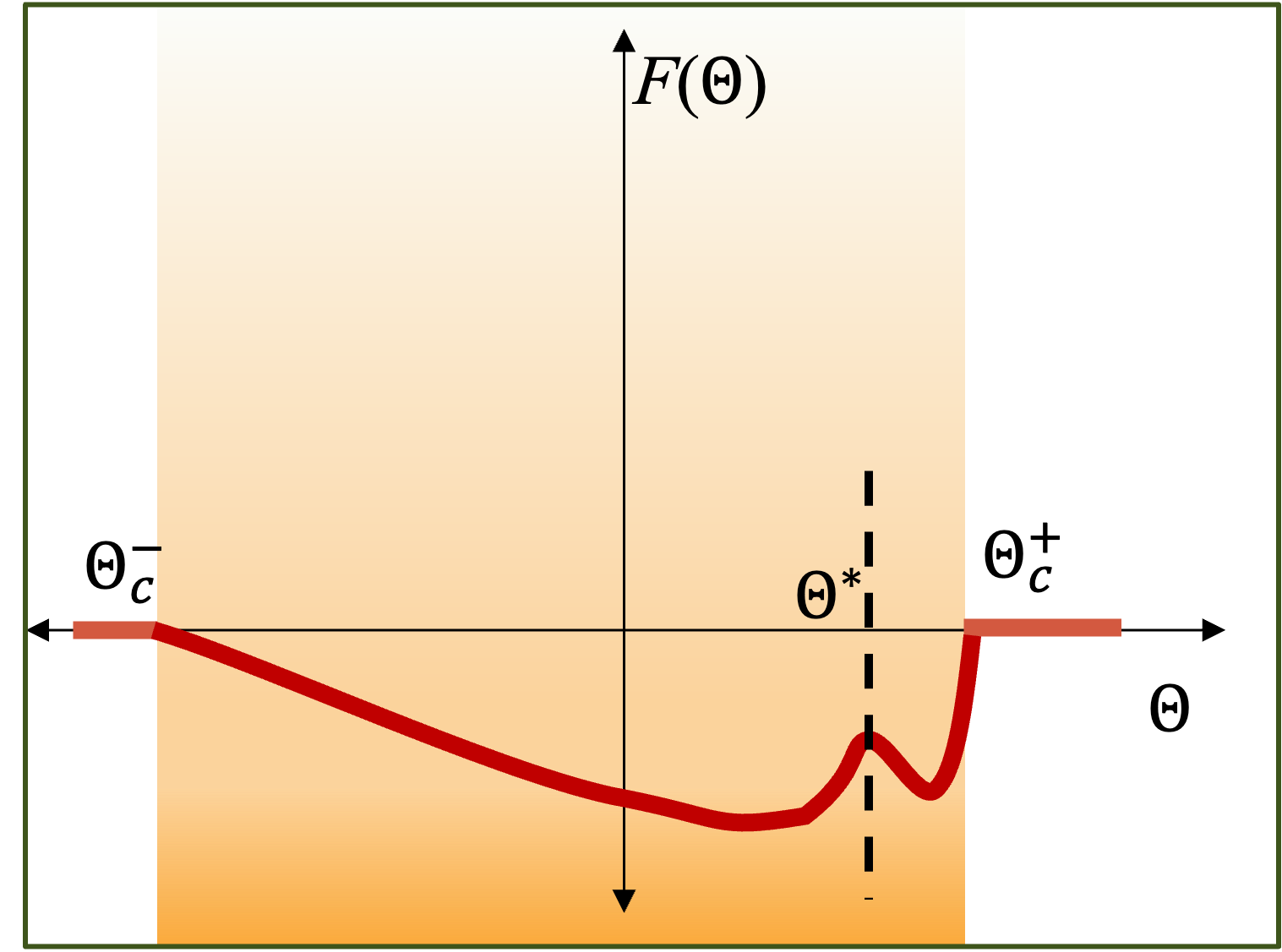}
\includegraphics[width=0.45\columnwidth]{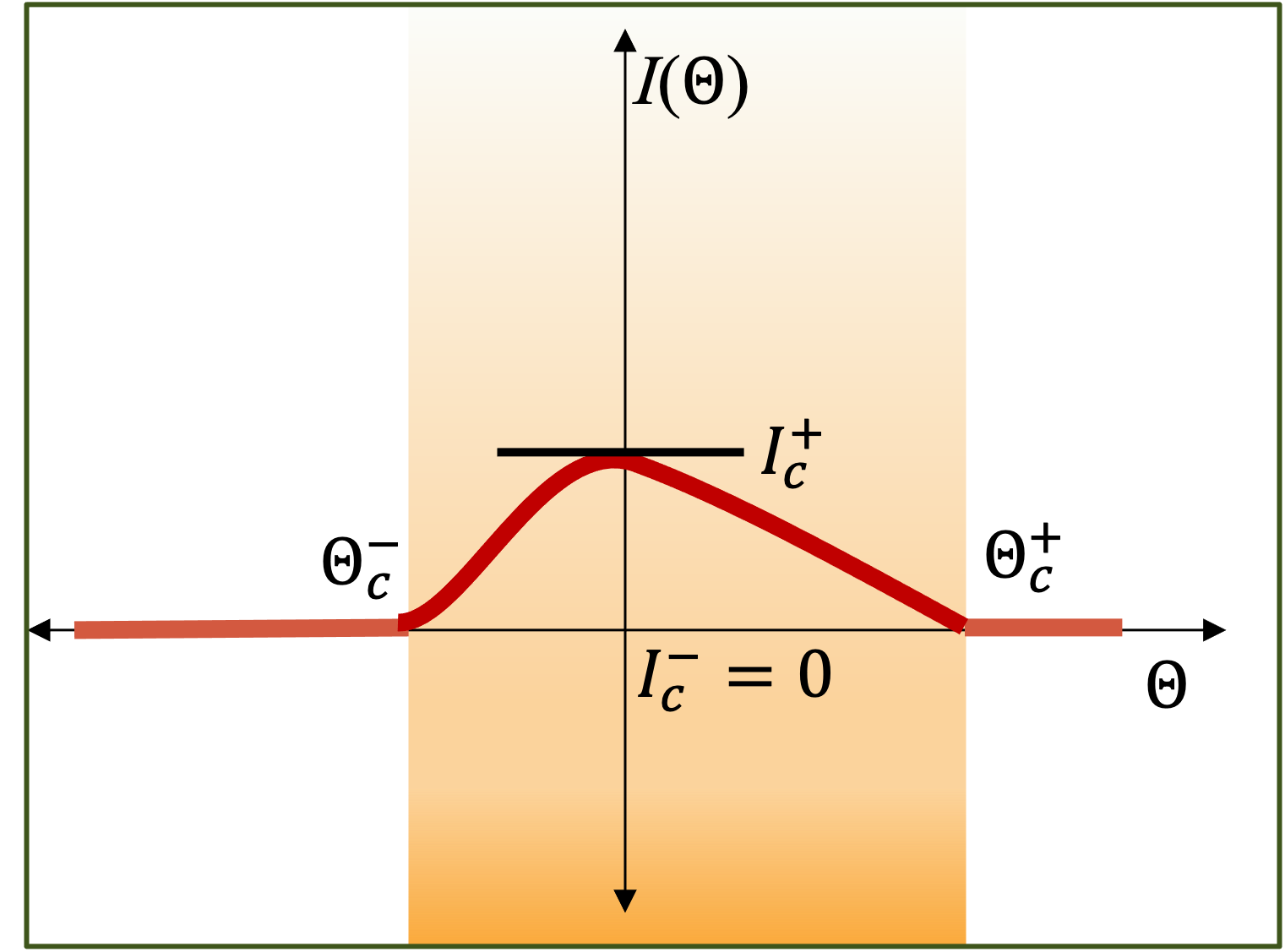}
\includegraphics[width=0.45\columnwidth]{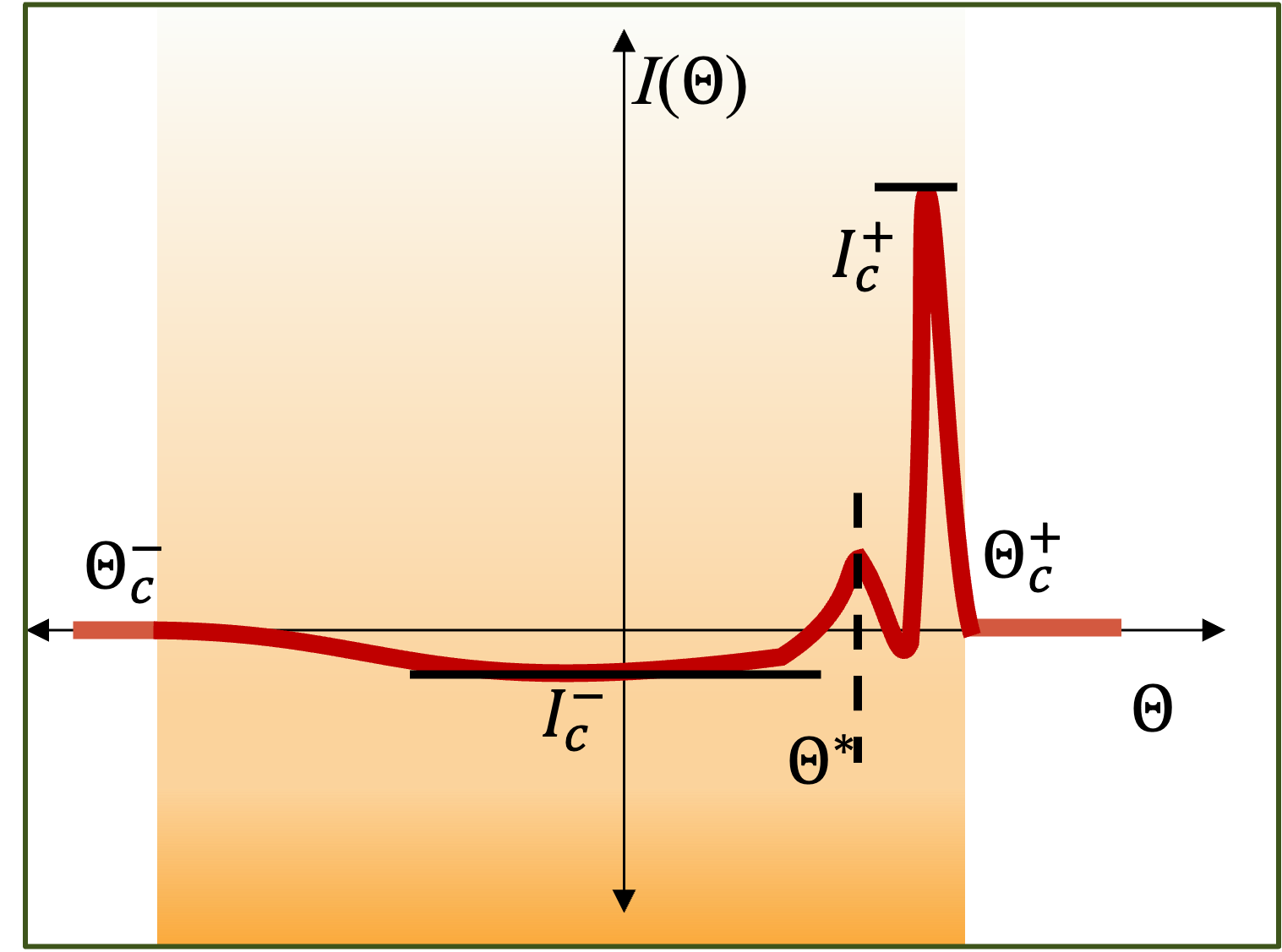}
\caption{A perfect diode without fine-tuning ($\epsilon=0$, left) requires $I(\Theta)\ge0$ and hence a discontinuity of $F(\Theta)$. Physical systems allow only $\epsilon\ll1$ (right), enabled by a broad region of small $I<0$ while continuity of $F(\Theta)$ is maintained by an internal non-analyticity at $\Theta^*$.
\label{fig:Schematic}}
\end{figure}

Suppose $I_c^- = 0$ and $I_c^+\neq0$. This implies $I(\Theta)\geq 0$ $\forall \Theta$ and $I(\Theta)>0$ strictly for at least some finite subinterval of $\Theta$ within the superconducting regime. This implies
\begin{equation}
	F(\Theta_c^+) - F(\Theta_c^-) = \frac{1}{2e} \int_{\Theta_c^-}^{\Theta_c^+} d\Theta\, I(\Theta) > 0
    \label{eq:F-diff}
\end{equation}
which contradicts \eqref{eq:Fqc}. Thus, $\epsilon=0$ forces $F(\Theta)$ to have a discontinuity; see Fig~\ref{fig:Schematic} (left). Such \emph{zeroth order} phase transitions occur in some systems that violate conventional thermodynamics via nonlocal interactions \citep{Maslov:2004aa}, absence of ergodicity \citep{Hou:2021aa}, or involvement of gravity \citep{Altamirano_2014,Altamirano2013,Altamirano2014,Wei2016,Frassino:2014aa,Zou2014,Jahani2013,Hennigar2015,Gunasekaran:2012aa,Kubiznak:2016aa,Dehyadegari2017,Dehyadegari2018}. In conventional systems that respect locality, $F(\Theta)$ must be continuous, so such phase transitions and hence $\epsilon=0$ are impossible.

Next, we ask,``can one obtain a perfect SD as a limit $\epsilon\to0$?" This limit is permitted by standard thermodynamics as it allows $I(\Theta)$ to have negative values $\propto\epsilon$ over a $\Theta$-interval $\propto1/\epsilon$ which can cancel the positive contributions in \eqref{eq:F-diff} and respect \eqref{eq:Fqc}. We now prove by contradiction that this geometric condition mandates a non-analyticity in $F(\Theta)$ as $\epsilon\to0$.

Suppose $F(\Theta)$ is analytic for $\Theta\in(\Theta_c^-,\Theta_c^+)$. Physically, this means a single condensate type -- say, $\Delta_1$ -- is the ground state for all $\Theta$ in \eqref{eq:F-eff}, but $F(\Theta)$ may have multiple local minima in $\Theta$ which define different finite-$q$ ground states for the Cooper pairs with the same internal structure. Now, consider the family of analytic functions parametrized by $\lambda$, $F_\lambda(\Theta)$, for $\Theta\in\left(\Theta_c^-(\lambda),\Theta_c^+(\lambda)\right)$. By \eqref{eq:Iq}, $I(\Theta)$ must be analytic as well and generically produce a diode asymmetry $\epsilon(\lambda)$. Now, suppose $\epsilon(\lambda)\to0$ as $\lambda$ tends to some value $\lambda_0$. Then, the negative segments of $I_\lambda(\Theta)$ vanish in the limit, i.e., $I_{\lambda_0}(\Theta)=0$ for finite segments of $\Theta$. By the identity theorem for analytic functions, such a function cannot be analytic unless $I_{\lambda_0}(\Theta)=0$ identically $\forall \Theta$. Thus, { we conclude that} $I_{\lambda_0}(\Theta)$ and hence, $F_{\lambda_0}(\Theta)$ must have non-analyticities in the interval $\Theta\in\left(\Theta_c^-(\lambda_0),\Theta_c^+(\lambda_0)\right)$, as illustrated in Fig.~\ref{fig:Schematic} (right). In particular, $F(\Theta)$ must be a piecewise function of $\Theta$ consisting of broad, nearly flat $F'(\Theta)<0$ regions alongside finite $F'(\Theta)>0$ ones. The transitions between the broad and finite regions are non-analytic points where either the dominant condensate $\Delta_i$ changes, or textures such as domain walls form between different $\Delta_i$ or different local minima from the same $\Delta_i$.

Two points must be noted. First, the broad regions still correspond to finite condensation energy, i.e., finite depth for local minima, as the softening extends over a $\Theta$-interval of diverging length $\propto1/\epsilon$. Second, when $\Theta$ belongs to a compact domain of length $2\pi$ or $4\pi$, a true diverging length is not possible. Rather, this analysis applies to the region $|\Theta|\gg1$ where flux quantization is immaterial, and the diverging length corresponds to $q=\Theta/L$ exceeding some characteristic microscopic momentum scale of that system.

Having derived a general no-go theorem, we now ask, ``how small can $\epsilon$ be in single-component superconductors -- phases captured by a single condensate $\Delta_1$ in \eqref{eq:F-eff} -- described by an analytic function $F(\Theta)$?" This is a pertinent question as real materials usually do not harbor multiple coexisting or competing superconducting tendencies. To answer this, we adopt the standard Landau prescription of writing $F(\Theta)$ as a finite-degree polynomial in $\Theta$ over the interval $\Theta\in(\Theta_c^-,\Theta_c^+)$. Polynomial functions obey strong rigidity constraints: a polynomial that is small over a finite subinterval cannot be arbitrarily large elsewhere on that interval. This idea is solidified in App.~\ref{app:remez} for non-compact $\Theta$, yielding a lower bound:
\begin{equation}
    \epsilon \ge \frac{1}{T_n\!\left(\frac{1+r}{1-r}\right)},
    \label{eq:eps-Cheby}
\end{equation}
where $T_n(x)=\cosh[n\cosh^{-1}(x)]$ is the Chebyshev polynomial of degree $n$ in the range $x>1$, and
\begin{equation}
    r = 
\frac{\text{range of } \Theta \text{ over which } I(\Theta)>0}
     {\text{total superconducting range } (\Theta_c^+-\Theta_c^-)} \in (0,1) .
\end{equation}
For example, if $F(\Theta)$ has a single minimum at $\Theta=\Theta_0\in(\Theta_c^-,\Theta_c^+)$ where $I(\Theta_0)=0$, then $r=(\Theta_0-\Theta_c^-)/(\Theta_c^+-\Theta_c^-)$. Conversely, \eqref{eq:eps-Cheby} implies a bound on $n$:
\begin{align}
    n \ge \frac{\cosh^{-1}(1/\epsilon)}{\cosh^{-1}(1+2r)} \gtrsim \log(2/\epsilon) \text{ for }\epsilon\ll1
\end{align}
where we have dropped an $r$-dependent $O(1)$ prefactor in front of $\log(2/\epsilon)$. Thus, achieving a SD effect with $\epsilon\ll1$ requires a polynomial of parametrically large degree that grows logarithmically with $1/\epsilon$. Physically, this corresponds to a Landau theory capable of describing a large number -- up to $\sim n/2$ -- of competing superconducting phases with distinct Cooper-pair momenta, each associated with a local minimum of $F(\Theta)$.

App.~\ref{sec:JD} derives an analogous bound for Josephson
junctions, with \(n\) now interpreted as the highest Fourier harmonic
retained in \(F(\Theta)\). For a current-phase relation containing only
harmonics up to order \(n\), we find
\begin{equation}
\epsilon \ge
\left[
T_{2n}\!\left(\sec\frac{\pi r}{2}\right)
\right]^{-1},\quad n\ge \frac{\cosh^{-1}(1/\epsilon)}{\cosh^{-1}(\sec \pi r/2)} 
\end{equation}
Thus, any finite harmonic truncation leaves the $\epsilon$ bounded away from zero. Ref.~\cite{Souto2022} implicitly used this condition to propose a Josephson diode with high -- but not perfect -- efficiency built from a large number of harmonics. Beyond strict truncation, non-smoothness generates additional Fourier tails that can readily weaken the finite-harmonic bound, and lower $\epsilon$ can be realized if the Fourier coefficients are properly fine-tuned.

\begin{figure}
\includegraphics[width=0.95\columnwidth]{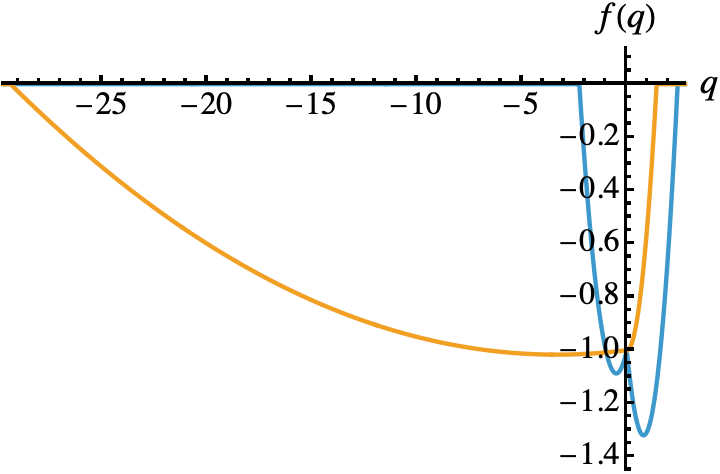}\\
\vspace{2mm}
\includegraphics[width=0.95\columnwidth]{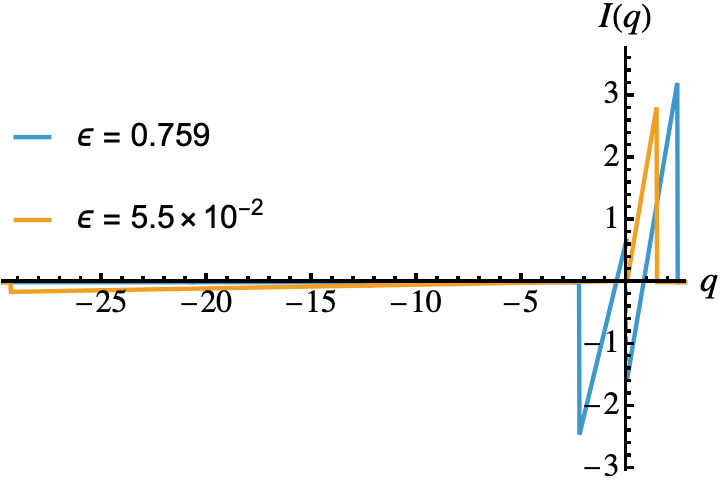}
\caption{
$f(q)$ and $I(q)$ for the Ising-coupled superconductor.
Blue curves show a generic finite-diode regime with parameters
$f_0=1$, $K_+=1$, $K_-=0.7$, $\gamma_+=0.8$, $\gamma_-=0.35$, yielding $f(q)$ minima at $q_*^+=0.8$, $q_*^-=-0.5$, $f(q)=0$ for $q>q_c^+\approx2.42$ and $q<q_c^-\approx-2.26$, and
$\epsilon=I_c^-/I_c^+\approx 0.759$.
Orange curves show a near-perfect diode regime with
$f_0=1$, $K_+=1$, $K_-=3\times10^{-3}$, $\gamma_+=\gamma_-=0.01$, yielding $q_*^+=0.01$, $q_*^-\approx-3.33$, $q_c^+\approx1.42$, $q_c^-\approx-29.4$ and
$\epsilon\approx 5.5\times10^{-2}$. Note, $|q_*^+|\ll|q_*^-|$ and $|q_c^+|\ll|q_c^-|$ in this case.
\label{fig:Ising}}
\end{figure}

\textit{Ising-coupled superconductor:} To concretize the above analysis, we consider an exactly solvable toy model with the necessary piecewise structure for $F(\Theta)$ which realizes the required softening of the $I(\Theta)<0$ branch. 

We work in terms of the phase twist wavevector $q=\Theta/L$. Specifically, consider a superconductor coupled to an Ising variable $\sigma=\pm1$, with free energy per unit length $f=F/L$ given by
\begin{equation}
{ {f}(q;\sigma)}=-f_0+\frac{K_\sigma}{2}\,q^2-\sigma\gamma_\sigma q.
\label{eq:IsingFq}
\end{equation}
where $K_\sigma>0$ are left/right stiffnesses and $\gamma_\sigma>0$ are linear magnetoelectric couplings. Minimizing over $\sigma$ at each
$q$ selects the lower branch and produces a piecewise condensation free energy
\begin{align}
f(q)=
\begin{cases}
-\!f_0+\dfrac{K_-}{2}q^2+\gamma_- q, & q<0,\\[4pt]
-\!f_0+\dfrac{K_+}{2}q^2-\gamma_+ q, & q>0,
\end{cases}
\label{eq:Fq_piecewise}
\end{align}
with a cusp at $q=0$ reflecting the switch of the internal Ising configuration. 
We assume the cusp is at $q=0$ for simplicity; in general, it can occur at non-zero $q$ and momentum must be measured relative to that point. Superconductivity exists when $f(q)<0$ or $q\in(q_c^-,q_c^+)$ where $q_c^\pm=\pm\frac{1}{K_\pm}\left(\gamma_\pm+\sqrt{\gamma_\pm^2+2K_\pm f_0}\right)$, and Cooper pairs in the two superconducting phases carry momenta $q_*^\pm=\pm\gamma_\pm/K_\pm$ where $f(q)$ has local minima. 

The corresponding supercurrent $I(q)$ is given by
\begin{equation}
I(q)=2e\,\partial_qf(q)=
\begin{cases}
2e(K_-q+\gamma_-), & q_c^-<q<0,\\[3pt]
2e(K_+q-\gamma_+), & 0<q<q_c^+,
\end{cases}
\label{eq:j_piecewise}
\end{equation}
$I(q)$ vanishes at $q_*^\pm$ and has a discontinuity of $2e(\gamma_++\gamma_-)$ at $q=0$. The conditions for a perfect SD are governed by the critical currents $I_c^\pm=\max[\pm I(0^\mp),\pm I(q_c^\pm)]$:
\begin{equation}
I_c^\pm = 2e\max\left(\gamma_\mp,\sqrt{\gamma_\pm^2+2K_\pm f_0}\right),
\label{eq:jc_from_Ising}
\end{equation}
Clearly,~\eqref{eq:jc_from_Ising} shows that $\epsilon=I_c^-/I_c^+\to0$ requires $\gamma_-$, $\gamma_+$ and $K_-$ to all vanish, so the $\sigma=-1$ branch becomes essentially $q$-independent, while the $\sigma=+1$ branch retains a finite stiffness $K_+$.

Fig.~\ref{fig:Ising} shows the behavior of this model for two representative parameter sets that yield finite and vanishing $\epsilon$. When all parameters are $O(1)$, corresponding to the blue curves, $f(q)$ has two clearly discernible potential wells at $q_*^\pm$ values of comparable magnitude, a cusp at $q=0$ and vanishes beyond positive and negative thresholds $q_c^\pm$ that are also comparable in magnitude. Moreover, $I(q)$ crosses zero at the well minima, and has finite positive and negative peaks at $q_c^\pm$ and on either side of the $q=0$ cusp. The depairing critical currents according to \eqref{eq:jc_from_Ising} also have similar magnitude: $I_c^+\approx3.21$, $I_c^-\approx-2.46$. On the other hand, when $\gamma_\pm$ and $K_-$ are $\ll1$ in units where $K_+=f_0=1$, as shown by the orange curves, the positive and negative wells as well as the extents of $f(q)$ have vastly different characteristic momenta: $|q_*^+|\ll |q_*^-|$, $|q_c^+|\ll |q_c^-|$. This, in turn, leads to $\epsilon\approx 5.5\times10^{-2}\ll1$.

{ \textit{Relevance to existing proposals and realizations:}
Our results constrain SD behavior only when the measured state corresponds to unconstrained minimization of an equilibrium free energy with respect to the transport-conjugate phase variable. Existing proposals and experiments that report high diode efficiency evade this setting in one or more of the following ways.

First are \emph{constrained equilibrium} routes, where the condensate phase is not free to relax because it is embedded in a larger superconducting environment, such as proximity-based platforms where a parent superconductor induces pairing in a metallic subsystem \citep{Hosur2023,Anh:2024aa} and prevents the latter from minimizing its own free energy. Second are \emph{dynamical stability}-based routes, where the observed response is controlled by a local minimum or by the loss of stability of a selected branch rather than by the global minimum. This includes intrinsic criticality-based proposals involving competing superconducting orders \cite{Shaffer2024,ChakrabortyBlackSchaffer2025,Yuan2022,Scammell_2022}, where large diode efficiency is obtained by analyzing a chosen minimum and its neighborhood. For example, Ref.~\cite{Yuan2022} notes that the SD effect vanishes if a $q>0$ condensate relaxes to the competing $q<0$ branch, while switching or domain-wall formation between the branches may intervene in practice. Current-biased Josephson devices also typically are in this category, since measured switching currents often probe escape from a metastable minimum in a tilted washboard landscape rather than extrema of an equilibrium current-phase relation. Third are \emph{genuine nonequilibrium} routes, where no static equilibrium free energy governs the measured state. This includes explicitly dissipative, driven, or time-dependent protocols, periodically driven systems, and other steady states sustained by continuous energy flow. Several Josephson diode proposals and realizations invoke such ingredients directly \cite{Daido2023,ShafferLi2025,Borgongino:2025aa}. Fourth, multiterminal devices \citep{Chiles:2023aa,Coraiola:2024aa} lie outside the purview of our two-terminal constraints and can achieve ideal efficiencies \cite{Chiles:2023aa} without violating them.

Lastly, not all large-efficiency proposals evade our assumptions. Our bounds explicitly allow high diode efficiency when the equilibrium free energy contains sufficiently many harmonics, yielding a highly structured current-phase relation. Josephson interferometric constructions realize this permitted route directly \citep{Souto2022,Haenel2022}.

\emph{Experimental implications:}
The above classification provides both explanatory and predictive guidance for interpreting present experiments. First, it clarifies why large $\eta$ has generally been easier to realize in Josephson platforms, which typically fall in one of the above categories, than in nominally intrinsic bulk superconductors, which more closely approximate the unconstrained equilibrium setting analyzed here. 
Second, the classification suggests a practical diagnostic for experiments. If a large diode effect is observed in a nominally intrinsic platform, in the absence of clear external phase constraints or explicit nonequilibrium drive, a natural interpretation is that transport either does not sample the global free energy landscape, for example due to metastability, or the free energy develops non-analytic structure, for instance, due to competing orders within the superconducting regime. These possibilities are pertinent to the near-perfect diode behavior reported in twisted trilayer graphene \cite{Lin:2022aa}. 
If interpreted within an unconstrained equilibrium description, this implies that other local minima as a function of $q$ must exist elsewhere in the superconducting landscape, separated by energetic or dynamical barriers. This picture lends itself to systematic falsification: $\epsilon$ is expected to be sensitive to protocols that facilitate relaxation across such barriers such as slower parameter ramps, thermal cycling, or added noise. Conversely, robustness against such probes would motivate closer examination of nonequilibrium mechanisms or competing orders that render the equilibrium free energy non-analytic. Thus, beyond bounding equilibrium diode response, our framework provides a route to distinguish between possible physical origins of high-efficiency diode behavior.

\medskip

In summary, this work shows that exact perfect equilibrium superconducting diodes are forbidden by continuity of the free energy, while near-perfect behavior requires singular restructuring of the superconducting landscape within an unconstrained equilibrium description or departure from global equilibrium. This explains why strong diode response is difficult to realize in simple intrinsic superconductors but common in engineered Josephson and driven platforms. Our work provides a unified thermodynamic lens for interpreting, comparing, and designing superconducting diodes.}

\section*{Acknowledgements}
We thank Steven Kivelson, Chandra Varma and T. Senthil for useful discussions. We are grateful to the Department of Physics and Astronomy at Rice University, the Max Planck Institute for the Physics of Complex Systems, and the Aspen Center for Physics, which is supported by National Science Foundation grant PHY-2210452, for their hospitality during different parts of this project. This research was supported by the Department of Energy Basic Energy Sciences grant no. DE-SC0022264.

\appendix

\section{Derivation of the Remez--Chebyshev bound}
\label{app:remez}

In this Appendix we derive the quantitative bound on the diode asymmetry
\(\epsilon=I_c^-/I_c^+\) quoted in the main text for finite-order Landau
theories. We assume that within a Landau expansion truncated at order
\(n\), the supercurrent
\begin{equation}
I(\Theta)=2e\,\partial_\Theta F(\Theta)
\end{equation}
is a real polynomial of degree \(\le n\) on the superconducting phase
window
\begin{equation}
\mathcal{I}=[\Theta_c^-,\Theta_c^+] .
\end{equation}

Using the definition of \(r\) as in the main text,
\begin{equation}
r \equiv
\frac{\text{length of region where } I(\Theta)>0}{|\mathcal{I}|}\in(0,1),
\end{equation}
let \(\mathcal{I}_+\subset\mathcal{I}\) denote the set where
\(I(\Theta)>0\), and \(\mathcal{I}_-\subset\mathcal{I}\) the set where
\(I(\Theta)<0\). These intervals need not be contiguous. Then
\begin{equation}
|\mathcal{I}_+|=r|\mathcal{I}|,\qquad
|\mathcal{I}_-|=(1-r)|\mathcal{I}|.
\end{equation}

The critical currents are
\begin{equation}
I_c^+=\sup_{\Theta\in\mathcal{I}} I(\Theta),\qquad
I_c^-=\sup_{\Theta\in\mathcal{I}}[-I(\Theta)],
\end{equation}
and \(\epsilon=I_c^-/I_c^+\). The strategy is to show that for given
\(I_c^-\) and \(n\), the positive critical current \(I_c^+\) cannot be
arbitrarily large, yielding a lower bound on \(\epsilon\).

To ensure superconductivity is absent outside the interval, we must
have \(I(\Theta)<0\) just to the right of \(\Theta_c^-\), and
\(I(\Theta)>0\) as \(\Theta\to \Theta_c^+\) from the left. We therefore
introduce the rescaled variable
\begin{equation}
x=a(\Theta-\Theta_c^-)-1,
\label{eq:affine}
\end{equation}
which maps the leftmost point of \(\mathcal{I}_-\) to \(-1\). We choose
\(a\) such that the subset \(\mathcal{I}_-\) maps to a (possibly disconnected) set \(E\) of total length
2. Under Eq.~\eqref{eq:affine}, the full superconducting window
\(\mathcal{I}\) maps to
\begin{equation}
\left[-1,\frac{1+r}{1-r}\right].
\end{equation}

Next define the normalized polynomial
\begin{equation}
p(x)=\frac{I(\Theta(x))}{I_c^-}.
\end{equation}
By construction, \(p(x)\) has degree \(\le n\) and satisfies
\begin{equation}
|p(x)|\le1,\qquad x\in E,
\end{equation}
since on \(\mathcal{I}_-\) one has \(-I(\Theta)\le I_c^-\).

The Remez inequality states that if a real polynomial \(p(x)\) of degree
\(\le n\) satisfies \(|p(x)|\le1\) over an interval
\(E\subset[-1,(1+r)/(1-r)]\) of length at least 2, then
\begin{equation}
\sup_{x\in\left[-1,\frac{1+r}{1-r}\right]}|p(x)|
\le
T_n\!\left(\frac{1+r}{1-r}\right),
\end{equation}
where \(T_n\) is the Chebyshev polynomial of the first kind.

Undoing the normalization gives
\begin{equation}
\frac{\sup_{\Theta\in\mathcal{I}}|I(\Theta)|}{I_c^-}
=
\frac{I_c^+}{I_c^-}
=
\frac{1}{\epsilon},
\end{equation}
and therefore
\begin{equation}
\epsilon
\ge
\frac{1}{T_n\!\left(\frac{1+r}{1-r}\right)}.
\label{eq:remez-bound}
\end{equation}

{ 
\section{Bound on equilibrium Josephson diodes}
\label{sec:JD}

In this Appendix we derive a quantitative bound on \(\epsilon\) for
Josephson systems whose current-phase relation contains only a finite
number of harmonics, and then show how this bound is rapidly weakened
when the Josephson energy is not fully smooth.

We consider a \(2\pi\)-periodic Josephson energy \(F(\Theta)\), with
supercurrent
\begin{equation}
I(\Theta)=2e\,\partial_\Theta F(\Theta).
\end{equation}
If the Fourier expansion of \(F\) is truncated at harmonic order \(n\),
then
\begin{equation}
F(\Theta)=F_0+\sum_{m=1}^{n} A_m \cos(m\Theta+\delta_m),
\end{equation}
and therefore
\begin{equation}
I(\Theta)=
-2e\sum_{m=1}^{n} mA_m \sin(m\Theta+\delta_m),
\end{equation}
which is a trigonometric polynomial of degree \(\le n\). The
equilibrium critical currents are
\begin{equation}
I_c^+=\sup_\Theta I(\Theta),\qquad
I_c^-=\sup_\Theta[-I(\Theta)].
\end{equation}

\subsection{Finite number of harmonics}

Let \(r\in(0,1)\) denote the fraction of the \(2\pi\) period over which
the current is positive,
\begin{equation}
r=
\frac{\text{range of }\Theta\text{ where } I(\Theta)>0}{2\pi}.
\end{equation}
The complementary region, of measure \(2\pi(1-r)\), is where
\(I(\Theta)\le0\). Define
\begin{equation}
p(\Theta)=\frac{I(\Theta)}{I_c^-}
\end{equation}
so that \(|p(\Theta)|\le1\) throughout the negative-current region.

Applying the sharp Remez inequality for trigonometric polynomials on the
circle \cite{tikhonov2018sharpremezinequality}, one obtains
\begin{equation}
\sup_{\Theta\in[0,2\pi)} |p(\Theta)|
\le
T_{2n}\!\left(\sec\frac{\pi r}{2}\right),
\end{equation}
where \(T_m\) is the Chebyshev polynomial of the first kind. Hence
\begin{equation}
\frac{I_c^+}{I_c^-}
=
\frac{1}{\epsilon}
\le
T_{2n}\!\left(\sec\frac{\pi r}{2}\right),
\end{equation}
or
\begin{equation}
\epsilon
\ge
\left[
T_{2n}\!\left(\sec\frac{\pi r}{2}\right)
\right]^{-1}.
\label{eq:JD-bound}
\end{equation}

This bound is finite for every finite \(n\). Thus, if the sign
structure of the current occupies a nonzero fraction \(r\) of the phase
circle, a finite number of Josephson harmonics cannot generate an
arbitrarily strong equilibrium diode effect.

Using \(T_m(\cosh\alpha)=\cosh(m\alpha)\), the bound may also be written
as
\begin{equation}
\epsilon\ge \operatorname{sech}[2n\,c(r)],
\end{equation}
with
\begin{equation}
c(r)=\operatorname{arccosh}\!\left(\sec\frac{\pi r}{2}\right).
\end{equation}
For large \(n\),
\begin{equation}
\epsilon \sim 2e^{-2n c(r)}.
\label{eq:JD-largeM}
\end{equation}

Thus, the practical lower bound decreases rapidly with harmonic number:
in principle, only a modest number of suitably tuned harmonics can
already produce very small \(\epsilon\).

\subsection{Including non-differentiability}

We now ask how such large effective harmonic order arises naturally when
the Josephson energy is not fully smooth.

Suppose the first non-smooth feature of \(F(\Theta)\) appears in its
\((k+1)\)-st derivative, i.e., \(F\in C^k\) but
\(F\notin C^{k+1}\). Repeated integration by parts gives the large-\(m\)
asymptotics
\begin{align}
|A_m|&\sim \frac{1}{m^{k+1}},\\
|I_m|&\sim \frac{1}{m^{k}},
\end{align}
up to model-dependent prefactors and possible logarithmic corrections.


To compare with the finite-harmonic result, decompose
\begin{equation}
I(\Theta)=I^{(\le n)}(\Theta)+I^{(>n)}(\Theta),
\end{equation}
where \(I^{(\le n)}\) contains modes \(m\le n\), while the tail
contains modes \(m>n\).

The low-harmonic sector obeys Eq.~\eqref{eq:JD-bound}, whereas the tail
satisfies
\begin{equation}
\|I^{(>n)}\|_\infty
\le
\sum_{m>n}|I_m|
\sim
\begin{cases}
n^{1-k}, & k>1,\\
\log n, & k=1,\\
n, & k=0.
\end{cases}
\end{equation}

Combining this with Eq.~\eqref{eq:JD-largeM} gives the schematic estimate
\begin{equation}
\epsilon \gtrsim C e^{-2n c(r)}-C' f_k(n),
\end{equation}
where
\begin{equation}
f_k(n)=
\begin{cases}
n^{1-k}, & k>1,\\
\log n, & k=1,\\
n, & k=0,
\end{cases}
\end{equation}
and \(C,C'>0\) depend on normalization and tail prefactors.

The key point is that the finite-harmonic bound is already exponentially
small, so even weak algebraic or logarithmic tails can readily
overwhelm it. Thus, modest non-smoothness can effectively trivialize
the finite-harmonic obstruction. 

\subsection{Remark on \(4\pi\) periodicity}

In systems hosting Majorana zero modes, the Josephson energy may become
\(4\pi\)-periodic:
\begin{equation}
F(\Theta+4\pi)=F(\Theta).
\end{equation}
The above reasoning carries through after redefining the fundamental
domain. The Fourier expansion then involves modes
\(e^{im\Theta/2}\) with integer \(m\), but the conclusions are
unchanged: finite harmonic content or sufficient smoothness prevents an
arbitrarily strong equilibrium diode effect.

\medskip

In summary, periodicity replaces interval geometry, but the central
principle remains the same: a perfect equilibrium Josephson diode
requires either unbounded harmonic complexity or sufficiently strong
nonanalyticity in the Josephson energy.}

\bibliography{library}

\end{document}